# The cosmological constant problem: from Newtonian cosmology to the greatest puzzle of modern theoretical cosmology

Federico Scali [ORCID 0009-0004-0637-561X]

University of Insubria, Department of Science and High Technology, Como, ITALY
And
INFN section Milan, ITALY.
`fscali@uninsubria.it`

**Abstract.** The cosmological constant problem is one of the greatest challenges in contemporary physics, since it is deeply rooted in the problematic interplay between quantum fields and gravity. The aim of this work is to review the key conceptual elements needed to formulate the problem and some ideas for a possible solution. I do so by weaving a *fil rouge* from Newtonian cosmology, through general relativity and the standard model of relativistic cosmology (Λ-CDM), up to the theory of quantum fields.

In the first part, the issues with the application of Newtonian gravity to an infinite and static universe are addressed, observing how a cosmological term in the Poisson equation would stabilize a homogeneous matter distribution. A toy derivation of the Friedman equations using only Newtonian arguments is also shown. In the second part, the conceptual path leading to general relativity and the Λ-CDM model is laid down, with particular emphasis to the historical introduction of the cosmological constant and its new role after the discovery of the accelerated expansion of the universe. Finally, the problem is formulated within the framework of quantum field theory. Its many facets are discussed together with the criticalities in the formulation and some of the leading ideas for its solution are outlined.

**Keywords:** Newtonian Cosmology, General Relativity, Relativistic Cosmology, Accelerated Expansion, Cosmological Constant Problem, Vacuum energy, Inhomogeneous Cosmologies.

# 1 Introduction

Since mankind has inhabited the Earth, the same view has been offered to anyone who has ever looked at the celestial vault in a very bright night: a great number of stars and galaxies distributed across the sky. And one begins to wonder, is this distribution homogeneous? Does it extend indefinitely in space, or is it finite? Does it change in time? During the course of history, the answers to these questions have accompanied radically different views of the world.

A philosopher of the archaic Greece would have referred to the universe using the word *kosmos*, which means order and harmony, as opposite to *chaos*[1] which is the lack of them. In fact, according to Greek cosmogony *kosmos* abruptly arose from *chaos*, and from that point onward the Earth and the stars came into being. In the medieval age, a scientist would have answered according to the Aristotelian-Ptolemaic system. They would have said that the universe is limited, both in time and space, and it is immutable. As for the stars, they would have argued that these are organized in concentric spheres and move around the Earth in a complex combination of cycles and epicycles.

Today we give very different answers to these millennial questions. Not only we have evidences that the universe extends indefinitely in space, but we also know that it is expanding. More than that, the observations of distant supernovae showed that the expansion gets faster as the time flows. All of this was discovered in the course of the last century and it was of revolutionary importance, since it forced a paradigmatic shift from an immutable *kosmos* to an evolving universe. But what is the cause of such a dynamics of the universe?

The answer of our cosmological model (the $\Lambda$-CDM) is to postulate the existence of a constant parameter, the *cosmological constant* $\Lambda$, which would be the "engine" responsible for the accelerated expansion. It was first introduced by Einstein as a fundamental quantity of nature characterizing the gravitational interaction, which would allow the existence of a static and finite universe. Today, a definitive verdict on the fundamental character of the cosmological constant still lacks, but it is generally thought as a sounder possibility that a deeper explanation is hiding.

For example, a natural motivation could be given in terms of entities even more fundamental than particles: the quantum fields.[2] These can interact gravitationally even in vacuum (i.e. without producing any particle), since they carry a minimum amount of energy, and would affect the dynamics of the universe in the same way of a cosmological constant. It would be natural, therefore, to identify the latter as the effect of the vacuum energy. Unfortunately, straightforward calculations result in an enormous discrepancy between the value predicted from the theory of quantum fields and the cosmological observations. In the following I will extensively address this point, which is known as the cosmological constant problem, and show how it is far less trivial than one may think. Its roots dig deep down to the interplay between quantum fields and gravity, and ultimately to the conditions which make our theories natural descriptions of the physical world.

---

[1] The word *chaos* derives from the verb *chásko*, which means to open the mouth widely and yawn. Therefore, according to Greek mythology "at the beginning of the world there was something like an immense open mouth, a huge yawn, empty and dark: the *chaos*. A profound and dark abyss ruled only by disorder" (Bettini 2015).

[2] The reason why a deeper motivation for the cosmological constant may be needed is subtle. Up to the discovery of the accelerated expansion the idea of a static universe lied at the foundations of the scientific paradigms; therefore, it came natural for Einstein to introduce an additional constant in the physical world to support such a fundamental fact. But for us this is not the case, since we do not conceive the current dynamics of the universe as fundamental. I should also mention that particle physics is not the only route to address the effect of the cosmological constant, as it will be discussed at the end of the paper.

The work is organized in three main sections. In the first the issues connected with the application of the Newtonian theory to an infinite universe are discussed. In the second, some key aspects of general relativity are reviewed and the main differences with a geometric formulation of Newtonian gravity are outlined. Moreover, the standard cosmological model is introduced, giving particular emphasis to the historical introduction of the cosmological constant and the discoveries concerning the expansion of the universe. In the final section, the cosmological constant problem is formulated within the framework of the quantum theory of fields. The problems associated with a straightforward calculation of the vacuum energy are discussed, together with the naturalness of the result and in comparison with the cosmological observations. Eventually, some of the leading ideas and perspectives towards a possible solution are discussed.

## 2 Newtonian Cosmology

This section is dedicated to Newtonian gravity and its application to the universe. This is problematic and leads to divergences in the gravitational field; nonetheless, a preliminary version of the Einstein equations in cosmology can be deduced only using Newtonian arguments. The resources for the section are mainly (Norton 1998) and (Einstein 1916b).

### 2.1 Newtonian gravity and the universe

Beginning with the second half of the sixteenth century, the Aristotelian-Ptolemaic system was gradually dismantled, in western thinking, during the century of the Copernican revolution (Kuhn 1957). The latter traditionally began with Copernicus' *De revolutionibus orbium coelestium,* in which a heliocentric system is proposed, and over one hundred years later it culminated with the majestic treatise of Newton, the *Philosophiae Naturalis Principia Mathematica* (Newton 1687), (Chandrasekhar 1995), in which the principles of classical mechanics and the universal law of gravitation are formulated.

Of this century or so of crisis, what is of interest for this section is the fact that, after Newton's Principia had been absorbed, the idea settled within the scientific community that the universe is infinite, both in time and space, and that there is no privileged spot to observe it (Copernican principle). Moreover, the distribution of matter would be immutable (i.e. static) and would extend homogeneously up to infinity.[3] In the following the difficulties arising when this view of the cosmos is embedded within the framework of classical mechanics are discussed.

---

[3] While it is easy to backtrack the roots of the Copernican principle in the process of the scientific revolution, it is not so for the idea of an infinite universe. Maybe, it was just easier to settle the issue of an observable boundary by pushing it to the unobservable infinity, instead of considering a space with no boundary at all.

According to Newtonian theory, the matter distribution is in equilibrium if the resulting of the gravitational forces acting upon every one of its elements is null. To compute it, let me first consider the gravitational field $g$ inside a uniform sphere of radius R and constant density $\rho(r) = \rho$. Using Gauss' law

$$g(r \leq R) = -\frac{4\pi G}{r^2} \int_0^r dr' r'^2 \rho(r') = -\frac{4\pi G}{3} \rho\, r. \tag{2.1}$$

Therefore, in the limit of an infinite, uniform matter distribution, $R \to \infty$, the force exerted on a point particle placed at infinity diverges. By symmetry, this is also the force exerted on a point mass at finite distance from the origin of the coordinates (up to a sign); hence, the distribution is unstable.

The first to lay down this problem in a mathematically precise fashion was the astronomer Seeliger in 1895 (Seeliger 1895), (Norton 1998); however, a surprisingly intuitive and clear derivation based on the concept of "lines of force" was given by Einstein in 1916 (Einstein 1916b).

There is another way to look at the same problem, from the point of view of the Newtonian potential. Recall that for a mass distribution $\rho$ the potential satisfies the Poisson equation

$$\nabla^2 \phi = 4\pi G \rho, \tag{2.2}$$

whose solution is determined up to a solution of the Laplace equation. The latter is necessarily trivial (i.e. everywhere constant) only if a constant boundary condition is imposed at infinity. But to have that, one should require $\rho \to 0$ faster than $\frac{1}{r^2}$, which is incompatible with an infinite and constant mass distribution.

What about a localized distribution, instead? Is it possible that we are simply immersed in a region of collapsed matter, in an otherwise empty space?

To answer, it is useful to think at such a system of stars and galaxies as a weakly interacting gas with a well definite internal energy. Every particle is subjected to a mean field potential binding it to the system; however, the statistical partition of the internal energy allows some of them to "jump over" the potential well and escape to infinity. As more particles leave this process becomes more unlikely; nonetheless, over a sufficient timescale it would be able to dilute the system, in clear contrast with the assumption of a static matter distribution.

At the turn of the nineteenth century, these observations led Seeliger and others to embark on the quest for a cosmological model compatible with classical mechanics. The proposals can be divided into two classes: those changing the hypotheses about the matter distribution;[4] and those formulating a large scale modification of the Newtonian gravitational law. Among the attempts falling in this second category, here I just mention the observation[5] Einstein places at the incipit of his famous cosmological considerations (Einstein 1917), since it paved the way to the introduction of the cosmological constant. Specifically, the German physicist considered a modified Poisson equation with the addition of a term proportional to the potential,

$$\nabla^2 \phi + \lambda \phi = 4\pi G \rho. \tag{2.3}$$

---

[4] In this respect, the hierarchical cosmology proposed by Charlier in 1908 deserves a mention (Norton 1998).

[5] Which the scientist says *it does not in itself claim to be taken seriously* (Einstein 1917).

The equation now admits the constant solution $\phi = \frac{4\pi G}{\lambda} \rho$; hence, the theory allows a constant mass distribution extending indefinitely in space.[6] At the end of the next section the reader will find a perfect similarity in the roles played by the cosmological term in the Poisson equation and the cosmological constant in the Einstein's equations.

## 2.2 A Newtonian look at the evolving universe

The former subsection dealt with the difficulties in the application of the Newtonian theory to a model of the universe filled with a static and uniform matter distribution. Here, just as an exercise let me apply the theory to a universe filled with a dynamical and uniform matter distribution.

Consider two elements of the matter fluid. Both the spherical symmetry and the homogeneity imply that only their distance $\Delta r \equiv |\vec{\Delta r}|$ may change in time, while the relative direction $\frac{\vec{\Delta r}}{\Delta r}$ should remain constant. Hence

$$\frac{d}{dt}\Delta r \equiv v_r = v_r(\Delta r, t). \tag{2.4}$$

Moreover, for sufficiently close elements

$$v_r(\Delta r, t) \simeq \frac{\partial v_r}{\partial \Delta r}\big|_{\Delta r=0}(t)\Delta r + o((\Delta r)^2), \tag{2.5}$$

where the zeroth order is absent, $v_r(0, t) = 0$. Defining $H(t) \equiv \frac{\partial v_r}{\partial \Delta r}\big|_{\Delta r=0}(t)$ and fixing the origin of the coordinates on one of the two elements, the position of the other evolves as[7]

$$\dot{\vec{r}}(t) = H(t)\vec{r}(t). \tag{2.6}$$

The vector $\vec{r}(t)$ can be expressed in terms of the function $a(t) \equiv e^{\int_{t_0}^{t} H(t')dt'}$ by separating the variables and integrating Eq. (2.6)

$$\vec{r}(t) = a(t)\vec{r}(t_0). \tag{2.7}$$

All of this was derived from purely kinematical considerations. But when the gravitational force comes into play, from the previous discussion it is clear that the boundary conditions on the matter distribution become a key factor, in order to have a well-defined gravitational field. For the moment, let me put aside this point and pretend that, no matter the extension of the source and the boundary conditions, the gravitational potential inside a spherical cavity is trivial.[8]

---

[6] Interestingly enough, this is the same term Newton considered in his revolving orbits theorem (Chandrasekhar 1995).

[7] The over dot indicates derivation with respect to time.

Again using Gauss' law, the total energy of an element of mass $m$ at radius $r$ is

$$E = \frac{1}{2}m\dot{r}^2 - \frac{4\pi G}{3}m\rho(t)r^2 = \frac{mr^2(t_0)}{a^2(t_0)}\left[\frac{\dot{a}(t)^2}{2} - \frac{4\pi G}{3}\rho(t)a^2(t)\right], \quad (2.8)$$

where $\rho(t)$ is the homogeneous distribution, which here is time dependent.
The conservation of energy thus implies

$$\left(\frac{\dot{a}(t)}{a(t)}\right)^2 - \frac{8\pi G}{3}\rho = \frac{K}{a^2(t)}, \quad (2.9)$$

with $K$ a real constant. This is a consistency equation that the function $a(t)$ must satisfy at every time.
Consider now the mass contained in a sphere of radius $r$

$$M = \frac{4\pi}{3}\left(\frac{a(t)}{a(t_0)}\right)^3 r(t_0)^3 \rho(t); \quad (2.10)$$

the classical principle of mass conservation implies

$$\dot{\rho}(t) + 3\frac{\dot{a}(t)}{a(t)}\rho(t) = 0. \quad (2.11)$$

Deriving once Eq. (2.9) and inserting Eq. (2.11), a dynamical equation for $a(t)$ is obtained

$$\frac{\ddot{a}(t)}{a(t)} = -\frac{4\pi G}{3}\rho. \quad (2.12)$$

The system of Eqs. (2.9) and (2.12) determines the small distance evolution of the cosmological fluid. To derive it one needs to get over the divergences in the gravitational field; hence, this should really be seen as a breakfast exercise. The reason why it is interesting will be clearer after the next section, when very similar equations will be obtained on the more solid ground of the general theory of relativity.

## 3 General relativity and cosmology

This section is devoted to the general theory of relativity and the standard cosmological model. The first part focuses on the conceptual path and the physical principles which led to Einstein's theory. The main resources for this part are the original works of Einstein (Einstein 1905abc), (Einstein 1907), (Einstein 1911), (Einstein 1915abcd), (Einstein 1916), (Lorentz et al 1923). The main differences with a geometric formulation of Newtonian gravity are also discussed (Cartan 1923, 1924), (Misner et al. 2017). The second half opens with a discussion of the observational and philosophical foundations of the cosmological principle (Ellis2007). Using the latter, the equations governing the dynamics of the universe are derived and Einstein's static solution, which historically led to the introduction of the cosmological constant, is presented. The discoveries of the expansion of the universe and of its current acceleration are eventually discussed. The main resources for the second part are (Ellis2007), (Weinberg 2008), (Misner et al 2017) and (Einstein 1917).

---

[8] In general relativity this is the result of the Birkhoff theorem (Weinberg 1972). The latter holds in spherical symmetry, no matter the extension of the source.

**3.1 General relativity**

Although general relativity is a theory of the gravitational field, the historical context in which it developed sees the electromagnetic field as the main character.
At the first lights of the nineteenth century, the experiments and the theoretical models developed by Young (Young 1804) and Fresnel (Fresnel 1816) definitely settled the idea that light propagates as waves. But in the Newtonian paradigm, which had proved extremely successful in the previous century, the propagation of waves could only be understood as the harmonic displacement of material points in a mechanical medium, just like sound waves in the air. Therefore, the newly discovered phenomena of interference and diffraction of light could fit within the traditional picture solely by admitting the existence of a fluid permeating space, amidst which light propagates: the luminiferous ether.
Now, if radiation propagates with velocity $c$ with respect to a frame at rest in the medium, it is understood that a moving observer should measure a different velocity of light, $\vec{c} - \vec{v}$ according to Galileo transformations. Such a difference would be an indirect indication of the medium, and most importantly, it should be detectable by means of terrestrial experiments since Earth is in non-inertial motion around the Sun.
This consideration boosted interest in experiments detecting a change in the speed of light, as the most famous Fizeau (Fizeau 1851) or Michelson-Morley (Michelson and Morley 1886) experiments, but all of them provided no evidence of anisotropies in the light propagation.
As it so often happens in history, these failed attempts did not acquire more relevance than a puzzling nuisance until a satisfactory theoretical explanation came into play. Towards the end of the century, the last formulation of Maxwell's electromagnetic theory (Maxwell 1865) and its strong success in accounting for the electromagnetic phenomena showed no need for a mechanical medium. This was the biggest hit for the ether conjecture[9] and made unavoidable the questions concerning a unification of the relativity principle with the new theory of light. The problem was that the Galileo transformations do not allow a mapping of the electric and the magnetic components among inertial frames which is compatible with Maxwell's equations; therefore, as long as the relativity principle was tied down to Galileo's transformations, there was no way to unify the former with the new theory.

---

[9] The appearance of Maxwell's theory, however, did not constitute the end of the ether. The conjecture was still incorporated in more articulated theories as the ether-dragging models (Whittaker 1989), in which the medium surrounding moving matter is partly dragged with it, and most importantly Lorentz's ether theory (Lorentz 1904), which envisaged many conceptual elements of special relativity.

As it is well known, the proposal which solved the conundrum came from Einstein in 1905 (Einstein 1905c). He observed that

> We have to take into account that all our judgments in which time plays a part are always judgments of simultaneous events. If, for instance, I say, "That train arrives here at 7 o'clock", I mean something like this: "The pointing of the small hand of my watch to 7 and the arrival of the train are simultaneous events". (Einstein 1905c).

As long as one is interested in events happening in the neighborhood of $A$ there is nothing wrong with this conflation, but for events happening at large distance from $A$ things become trickier: there is in principle no way to compare the readings of the clocks, unless these have been synchronized.[10] The fact that the synchronization of clocks at different places in space requires a nontrivial physical procedure is the first key conceptual element introduced by Einstein.

The second element was the acknowledgement that the relativity principle can be unified with Maxwell's theory if the speed of light is erected to a constant of nature, with a determined value no matter the state of (inertial) motion of the emitting body. This postulate clearly broke down the idea of a mechanical medium and the intuitive, Galilean composition of velocities among inertial frames. More than that, it marked an irreconcilable fracture with the Newtonian concepts of space and time.

> So we see that we cannot attach any absolute signification to the concept of simultaneity, but that two events which, viewed from a system of co-ordinates, are simultaneous, can no longer be looked upon as simultaneous events when envisaged from a system which is in motion relatively to that system. (Einstein 1905c).

It did not take too long before the German physicist began to wonder about an extension of the relativity principle to frames in accelerated motion with respect to each other (Einstein 1907). He was struck by a revelatory thought experiment which deserves a mention.

Assume an observer standing closed in an elevator, directed along the $z$ axis, in which they can perform any kind of experiment. Consider the two following possibilities: that the elevator is moving with constant acceleration $g$ in the positive $z$ direction; or that the elevator is held fixed in a uniform gravitational field $\phi$ so that $-g\hat{z} = -\vec{\nabla}\phi$. Is there any experiment they can perform to distinguish the two situations?

Intuition and experience suggest that the answer is in the negative, so that the two systems should be considered in all respects as physically equivalent. Hence, not only the inertial and the gravitational mass coincide, but all the physical laws have the same meaning for an observer in uniform acceleration, or under the effect of a uniform gravitational field.[11] The latter has no ontological existence, as the difference between the electric and the magnetic fields.

---

[10] Einstein's proposal for the synchronization is a natural one: consider a light ray departing from $A$ at $t_A$, arriving in $B$ at $t_B$ and again reflected to $A$ at $t'_A$. If the time elapsed to reach $B$, $t_B - t_A$, is equal to the time elapsed to bounce back to $A$ from $B$, $t'_A - t_B$, then the two clocks are synchronized. More generally, if the reading of the $B$ clock can be adjusted so that $t_B = \frac{1}{2}(t'_A + t_A)$, then the synchronization is possible and a common time can be defined. This procedure is not the only one possible, see (Macdonald 1983). Moreover, it is not mandatory to use a global time coordinate corresponding to the common reading of synchronized clocks. One should just be careful that for a generic time coordinate, the spatial surfaces of constant time do not necessarily correspond to surfaces of simultaneous events in space.

This observation extended the principle of relativity to frames in uniform acceleration, which is compensated by the appearance of gravity. Even more, the same principle extended to any local observer in any regular gravitational field; since, for sufficiently small regions of space and short time intervals, the variation of the field can be neglected.

The birth of general relativity had to wait the intense development of those ideas firstly appeared in 1907. But finally, in November 1915 Einstein published four papers (Einstein 1915abcd) containing the foundations of the general theory. Here I will refer to the more systematic exposition appeared afterwards (Einstein 1916), in 1916, which constitutes a final consolidation of those works.

From the previous discussion, there appear two founding principles for the new theory. The first is that "The laws of physics must be of such a nature that they apply to systems of reference in any kind of motion" (Einstein 1916). This would be called the general principle of relativity, as opposed to the special principle of relativity holding for systems in relative uniform motion. The second principle is that "The general laws of nature are to be expressed by equations which hold good for all systems of co-ordinates, that is, are co-variant with respect to any substitutions whatever (generally co-variant)" (Einstein 1916). The difference between the two may seem subtle but it is of key importance. On the one hand, the frame-invariance principle informs that the laws of nature have the same meaning for any observer, no matter their state of motion. Hence, for example, the Newtonian concept of force cannot enter the theory in a fundamental way, because it has an absolute meaning only for inertial observers (apparent, frame-dependent forces appear for non-inertial observers). On the other hand, the principle of general covariance is a statement concerning the coordinates: since there is no natural way to lay down a frame in finite portions of spacetime, the laws of physics should be written in forms which treat any system on equal footing. Because of the equivalence principle, the theory resulting from these two founding ideas is necessary a theory of the gravitational field.

It is not my intention to derive here all the technical facets of general relativity, see (Weinberg 1972) and (Misner et al 2017) for complete expositions. Here I just want to mention some key results which are useful to understand it and will be recurrent.

As noted by Minkowski (Minkowksi 1908), in special relativity the values of spatial lengths and time intervals depend on the reference frame, but the "displacement" (or line element) $ds$ of two infinitesimally near events in Minkowski space

$$ds^2 = -c^2 dt^2 + |\vec{d}x|^2 \equiv \eta_{\mu\nu} dx^\mu dx^\nu, \qquad (3.1)$$

---

[11] Another equivalent, maybe more familiar, way to express this concept is by saying that an inertial reference frame cannot be distinguished from a frame in free fall in a uniform gravitational field. This is sometimes called the Einstein's equivalence principle (Liberati 2015).

does not. In the formula, the four-components differential is $dx^\mu = (cdt, \vec{dx})$, hence the indices $\mu$ and $\nu$ run from 0 to 3,[12] and the coefficients $\eta_{\mu\nu}$ form the Minkowski metric,

$$\eta_{\mu\nu} = \begin{pmatrix} -1 & 0 & 0 & 0 \\ 0 & 1 & 0 & 0 \\ 0 & 0 & 1 & 0 \\ 0 & 0 & 0 & 1 \end{pmatrix}, \quad (3.2)$$

which is also invariant under a Lorentz transformation.[13]

Let me consider a general gravitational field in spacetime. Because of the locality of the equivalence principle, in the neighbourhood of every point $P$ there always exists a Lorentz frame $x_0^\mu$ in which the laws of special relativity apply as they are (Misner et al 2017). In a different and arbitrary coordinate system $x^\mu$, defined in a finite region of spacetime encompassing $P$, $ds$ reads

$$ds^2 = \eta_{\mu\nu} dx_0^\mu dx_0^\nu = \eta_{\mu\nu} \frac{\partial x_0^\mu}{\partial x^\rho} dx^\rho \frac{\partial x_0^\nu}{\partial x^\sigma} dx^\sigma \equiv g_{\rho\sigma} dx^\rho dx^\sigma, \quad (3.3)$$

where a general metric has been defined

$$g_{\rho\sigma}(x) = \eta_{\mu\nu} \frac{\partial x_0^\mu}{\partial x^\rho} \frac{\partial x_0^\nu}{\partial x^\sigma}. \quad (3.4)$$

Although Eqs. (3.3) and (3.4) hold in a neighbourhood of $P$, the metric $g_{\rho\sigma}(x)$ defines distances and time intervals in all the region covered by the system $x^\mu$. In this regard it should be stressed that Eqs. like (3.4) hold at every point in the domain of $x^\mu$ with different Lorentz systems; however, in the presence of a nonuniform gravitational field it is not possible to coherently join the latter in a single frame. This is expression of the locality of the equivalence principle. Also, notice how the dependence of $g_{\rho\sigma}$ from the point in spacetime represents the variability of the gravitational field in the new global reference frame.

The last point becomes even clearer by looking at the trajectory of a "freely falling" particle.[14] An observer falling with the particle would describe its (infinitesimal) trajectory as the solution of

$$\frac{d^2 x_0^\mu(\lambda)}{d\lambda^2} = 0. \quad (3.5)$$

If the particle moves from the spacetime event $A$ to $B$, the solution extremizes the invariant line between the two events

$$\delta_{x_0(\lambda)} \int_{\gamma_{AB}} ds(x_0(\lambda)) = 0, \quad (3.6)$$

---

[12] The summation over repeated indices is left intended, if not stated otherwise.

[13] In fact, the Lorentz transformations can be defined as the group of transformations leaving $\eta$ invariant (Tung 1985).

[14] A body is freely falling if its motion is determined only by the gravitational field.

where $ds$ is defined by means of the Minkowski metric in the Lorentz frame and $\gamma_{AB}$ is a path connecting $A$ to $B$. Because of general covariance, the same equation (3.6) holds in any other coordinate system

$$\delta_{x(\lambda)} \int_{\gamma_{AB}} ds(x(\lambda)) = 0, \tag{3.7}$$

but the element $ds$ is now defined using the metric $g_{\mu\nu}(x)$. Performing the variation one obtains the geodesic equation[15]

$$\frac{d^2 x^\mu}{d\lambda^2} + \Gamma^\mu_{\nu\rho} \frac{dx^\nu}{d\lambda} \frac{dx^\rho}{d\lambda} = 0, \tag{3.8}$$

where $\Gamma^\mu_{\mu\rho}$ are the Christoffel symbols[16] $\left(\partial_\mu \equiv \frac{\partial}{\partial x^\mu}\right)$

$$\Gamma^\mu_{\nu\rho} = \tfrac{1}{2} g^{\mu\sigma}(\partial_\nu g_{\rho\sigma} + \partial_\rho g_{\nu\sigma} - \partial_\sigma g_{\nu\rho}). \tag{3.9}$$

From Eqs. (3.8) and (3.9) it is evident that the deviation from a straight line is caused by $\Gamma^\mu_{\nu\rho}$, which depend on the derivatives of the metric. This is the effect of the gravitational field appearing in an accelerated frame; hence, it is natural to interpret $g_{\mu\nu}$ as a generalization of the gravitational potential and $\Gamma^\mu_{\nu\rho}$ as the corresponding gravitational field.

The last ingredient needed to complete the theory is the set of equations linking $g_{\mu\nu}$ to the distribution of bodies in spacetime (Einstein 1915b). See also (Weinberg 2008) and (Misner et al 2017) for a more contemporary approach.

Today, it is customary to write the Einstein equations as

$$R_{\mu\nu} - \tfrac{1}{2} g_{\mu\nu} R = \tfrac{8\pi G}{c^4} T_{\mu\nu}. \tag{3.10}$$

The right-hand-side is the matter sector,[17] described by the energy-momentum tensor $T_{\mu\nu}$, while the left-hand side is the "geometrical" one and retains the information about the spacetime metric.

Both the Ricci tensor $R_{\mu\nu}$ and the curvature scalar $R$ are obtained from "contractions" of the Riemann tensor

$$R^\mu_{\nu\rho\sigma} = \partial_\rho \Gamma^\mu_{\nu\sigma} - \partial_\sigma \Gamma^\mu_{\nu\rho} + \Gamma^\mu_{\rho\lambda} \Gamma^\lambda_{\nu\sigma} - \Gamma^\mu_{\sigma\lambda} \Gamma^\lambda_{\nu\rho}, \quad R_{\mu\nu} \equiv R^\rho_{\mu\rho\nu}, \quad R \equiv g^{\mu\nu} R_{\mu\nu}; \tag{3.11}$$

---

[15] Here I am assuming to take an affine parametrization for the trajectory; for a general parametrization the result is slightly different, see (Poisson 2004).

[16] The tensor $g^{\mu\nu}$ is the inverse of the metric, defined by $g^{\mu\rho} g_{\rho\nu} = \delta^\mu_\nu$.

[17] With matter I intend, hereinafter, every energy contribution from fields other than the gravitational one. Hence also, for example, from the electromagnetic field.

hence, Eq. (3.10) is a set of second order, nonlinear, partial differential equations in the ten components of the metric.

Physically, the Riemann tensor quantifies the attraction of two test particles freely falling next to each other, due to the variability of the gravitational field. If there exists a reference frame in which gravity is totally absent, then two neighbouring geodesics are straight lines in Euclidean sense, never crossing each other, and the Riemann tensor vanishes in that as in any other reference system.

The latter has also a definite geometrical meaning (Nakahara 2003). Imagine an observer at a point $A$ in spacetime endowed with a ruler. Consider them performing a round trip from $A$ to $A$, keeping the ruler always parallel to itself at every step of the way. In the presence of a general gravitational field the final orientation of the ruler will differ from the starting one, the difference being proportional to the Riemann tensor. Therefore, as a measure of the deviation from what would happen in Euclidean space, the Riemann quantifies the curvature of spacetime. In this respect, the Einstein equations tell how the curvature is sourced by the energy configuration of matter (Misner et al 2017).

This conclude the survey of the conceptual elements and the few technical tools needed for subsection 3.3, in which the machinery of general relativity will be applied to the whole universe. From this point onward, formulas are expressed in natural units (Misner et al 2017) $\frac{h}{2\pi} = c = 1$, where $h$ is the Planck constant.

## 3.2 The geometric formulation of Newtonian gravity

Before tackling cosmology, it is worth pausing for a moment and ask what the main differences between general relativity and Newtonian gravity are.[18] At first glance the answer seems easy: one is a geometric theory, in which the role of the gravitational potential is played by the metric on a spacetime manifold, while the other simply is not. However, things are subtler than this because the geometric formulation is not as fundamental as one may think, as it was shown by Cartan (Cartan 1923, 1924) shortly after the birth of general relativity.

To understand this point, consider the classical equation of motion of a particle in a gravitational field[19]

$$\ddot{x}_i(t) + \frac{\partial \phi}{\partial x_i} = 0, \qquad (3.12)$$

where the dot stands for derivation with respect to the Newtonian absolute time $t$. From a geometric point of view, the latter is a scalar field (clock field) defined a priori throughout all spacetime, which, because of the nature of the "Newtonian spacetime", induces a natural foliation in Euclidean 3D spacelike hypersurfaces ($t = constant$).[20] This latter structure is the Newtonian absolute space.

---

[18] Letting aside the experimental confirmations of the predictions of general relativity, see (Will 2014).

[19] From now on, Latin indices (i,j,k…) indicate the spacelike components and run from 1 to 3.

[20] As a consequence, on every surface of constant $t$ there exist a global coordinate system in which the spatial metric is the Euclidean metric $\delta_{ij}$.

Adopting $t$ as time coordinate ($x_0 = t$) and the Euclidean frame on every surface (Galilean frame), comparison with the geodesic equation (3.8) immediately gives the Christoffel symbols

$$\Gamma^i_{00} = \frac{\partial \phi}{\partial x_i}, \text{ all the other } \Gamma^\mu_{\nu\rho} \text{ vanish.} \tag{3.13}$$

Also, using the expressions (3.11) one can readily compute the curvature tensors

$$R^i_{0j0} = -R^i_{00j} = \frac{\partial^2 \phi}{\partial x_i \partial x_j}, \text{ all other } R^\mu_{\nu\rho\sigma} \text{ vanish,} \tag{3.14}$$

$$R_{00} = \frac{\partial^2 \phi}{\partial x_i \partial x_i}, \text{ all other } R_{\mu\nu} \text{ vanish,} \tag{3.15}$$

and write the corresponding "Newton field equations", with the mass density as the source term

$$R_{00} = 4\pi G \rho, \text{ all other } R_{\mu\nu} \text{ vanish.} \tag{3.16}$$

Written in this form, the field equations of Newtonian gravity are closer to those of general relativity with two main differences. The first is that this is still an equation in one degree of freedom, the Newtonian potential, while Eq. (3.10) is an equation in the ten components of the metric.[21] The second is that, in this form, the equation is not generally covariant, since it is explicitly written within the class of the Galilean frames. However, using the independent existence of the clock field it is easy to generalize the content of the equation to a coordinate independent statement (Misner et al. 2017)

$$R_{\mu\nu} = 4\pi G \rho \langle dt, e_\mu \rangle \langle dt, e_\nu \rangle, \tag{3.17}$$

where $dt$ is the one form associated to the clock field, $e_\mu$ is a coordinate basis unit vector and $\langle \cdot, \cdot \rangle$ is their inner product (Nakahara 2003).
In the same spirit it is possible to formulate the existence of an absolute time and an absolute space, with the associated preferred class of Galilean frames, and of a scalar gravitational potential by means of coordinate free axioms (Misner et al. 2017). From the latter, Newtonian gravity follows as a geometric theory.
The bottom line is that the difference between general relativity and classical gravity does not reside in the geometric formulation of the first, since also the second can be formulated in the same language. Instead, the difference lies in the a priori geometrical structures: a spacetime metric for general relativity, and a global time function, foliating spacetime in Euclidean surfaces for Newtonian gravity.[22]

---

[21] It can be shown that it is not possible to define a nondegenerate spacetime metric in Newtonian gravity (Misner et al. 2017).

[22] In this sense, from the point of view of modern physics (and purely at a theoretical level) a reason to prefer general relativity is simply because its geometrical structure does not select a priori any preferred reference frame. Consequently, the laws of physics have a clear meaning no matter the coordinate system. From case to case the symmetries of a specific problem can suggest a natural reference frame, but this is limited to the particular system under study.

**3.3 Cosmological principle and spacetime geometry**

The standard model of modern cosmology (Weinberg 2008), (Piattella 2018) resides on a powerful symmetry hypothesis upon our universe: the so-called cosmological principle. This states that at sufficiently large scales ($> 100$Mpc) the universe is spatially homogeneous, meaning that there are no preferred points, and isotropic, meaning that there are no preferred directions. In particular the principle implies that, at any point in space, a typical freely falling observer (like we are) would see the same spherically symmetric universe.[23] In fact, the cosmological principle could be equally understood as affirming the complete equivalence of such freely falling reference frames, one at each spatial point, which can transform to each other leaving invariant all the physical quantities of cosmological interest (Weinberg 1972).[24]

Before addressing the implications of the cosmological principle, it is important to break it down and understand what physical and philosophical motivations lie behind it. As for the isotropy, the astronomical observations accurately establish a very high degree of spherical symmetry about our location. For example, measurements of the cosmic microwave background (CMB) (Weinberg 2008) show in every direction a black body spectrum at the same mean temperature, with a precision beyond one part in $10^4$ once the dipole anisotropy[25] has been removed. On the other hand, arguing for homogeneity at the sole observational level is way more difficult (Ellis 2007), because we access spacetime only by our location (or its surroundings) and the observations are limited either by technology or by some unknown physics needed to interpret the results. For this reason, the main accepted motivation is rather a philosophical posture, essentially affirming that no peculiarity should be assigned to the location of any observer with respect to the rest of the universe (Ellis 2007). This is known as the Copernican principle.[26] When the latter is interpreted as the complete equivalence of freely falling observers at any spatial location, in the sense stated above, together with the exact isotropy at our location it implies the exact homogeneity of space.[27]

---

[23] A typical freely falling observer is an observer essentially evolving with the dynamics of the universe, without major effects from local gravitational fields. A futuristic astronaut whizzing from galaxy to galaxy would not see the universe as isotropic, not even over a coarse grained look: a stream of matter would flow against them, dictating a preferred direction. Specifying the observers is thus important in this case.

[24] From this, the invariance under spatial translations of every cosmological tensor follows (Weinberg 1972).

[25] In the standard picture, this is caused by the peculiar motion of the Earth with respect to the typical freely falling frames (Weinberg 2008).

[26] It was born within the Copernican revolution, referring to the paradigmatic shift of Earth's position in the Solar system, but with the birth of modern cosmology it was quickly adapted to the bigger picture of Earth's position in the universe.

[27] This can be understood quite intuitively. For any two points $A, B$ in space it is always possible to find a point $C$ at equal spatial distance from them. Since by hypothesis the universe is isotropic at $C$, the points $A, B$ must be equivalent. For a more rigorous derivation see (Weinberg 1972).

Also, it is worth to mention that, in the 1950's, the so-called steady state cosmology proposed by Bondi and Gold (Bondi and Gold 1948), and separately by Hoyle (Hoyle 1948) received

In such interpretation the consequences of the Copernican principle really are far reaching: a smooth maximal symmetry is imposed on the spatial hypersurfaces, ensuring the invariance of all the cosmological observations performed from any point. However, it should be kept in mind that the homogeneity is not an exact symmetry of the matter distribution: observations show structures up to ∼ 100 Mpc, with galaxy clusters and super clusters organized in thin filaments surrounding large void bubbles.[28] Homogeneity is thus recovered only by coarse graining over scales much larger than 100 Mpc. It follows that in using such strong interpretation of the Copernican principle two additional, crucial elements are assumed, which are not strictly implied by the formulation. The first is that the inhomogeneities should not appreciably affect the overall dynamics of the universe, so that the matter distribution can just be "smoothed out" over space. The second is that the reference frame of typical observers coincide with the coarse grained reference frame defined by the cosmological principle, so that our measured quantities are in one-to-one correspondence with the averaged ones.

These two sensible points will be central in motivating the inhomogeneous cosmologies discussed at the end of the paper; for the moment, the cosmological principle will be assumed in full without further discussion. As it will become evident in the following, the latter dramatically constrains the degrees of freedom of the metric (as, in fact, of every cosmological tensor), allowing a particularly simple treatment of the Einstein equations and an efficient use of the limited observational data. Nonetheless, the resulting standard model presents serious criticalities (Perivolaropoulos and Skar 2022), (Abdalla et al. 2022) and understanding whether the cosmological hypothesis could be responsible for them is currently an active field of research (Aluri et al 2023).

But what exactly implies the cosmological principle on the form of the gravitational field and the energy-momentum tensor?

As for the latter, the isotropy implies the perfect fluid form (Misner et al 2017)

$$T_{\mu\nu} = (p + \rho)u_\mu u_\nu + p g_{\mu\nu}, \tag{3.18}$$

where $u^\mu$ is the velocity field, $\rho$ is the energy density and $p$ is the pressure. The homogeneity instead forces both $\rho$ and $p$ to be functions of time only.

For the gravitational field, the symmetries single out three possibilities for the geometry of the three dimensional space (Weinberg 1972): the Euclidean plane; the 3-sphere; and the hyperbolic 3-space. Using as coordinate time the common notion of proper time for the isotropic observers, the invariant line element for the cosmological spacetime can be written as

$$ds^2 = -dt^2 + a^2(t)\left[\frac{dr^2}{1-kr^2} + r^2(d\theta^2 + \sin^2(\theta)\,d\varphi^2)\right], \tag{3.19}$$

---

attention. It was based on an enhanced cosmological hypothesis, or perfect cosmological principle, which besides spatial homogeneity and isotropy assumed that the universe always appears the same to the fundamental observers (homogeneity in time). This idea essentially faded once the CMB was discovered (Weinberg 2008).

[28] This is known as the cosmic web and appears to be the end of the hierarchical clustering at present time (Coil 2013).

with $k$ taking the values 1, 0, -1 respectively in the spherical, flat and hyperbolic case. This is the Friedman-Lemaître-Robertson-Walker (FLRW) metric, deduced for the first time in this form by Lemaître (Lemaître 1931). The positive definite function $a(t)$ is a scale factor, which regulates how the spatial distances vary with time. It is the only free function of the gravitational field and should be determined as a solution the Einstein equations. Notice that, with the convention of (Weinberg 2008) adopted here, the radial coordinate is dimensionless while the scale factor has dimensions of a length.[29]

Thanks to the cosmological principle, the Einstein equations take the particularly simple form

$$\begin{cases} \frac{\ddot{a}(t)}{a(t)} = -\frac{4\pi G}{3}[\rho(t) + 3p(t)], \\ \left(\frac{\dot{a}(t)}{a(t)}\right)^2 + \frac{k}{a^2(t)} = \frac{8\pi G}{3}\rho(t), \end{cases} \quad (3.20)$$

known as the system of the FLRW equations (Friedman 1922).[30] The functions $\rho$ and $p$ represent the density and pressure of all the matter fluids in the universe and are unknowns as well. However, the Friedman equations alone can determine only one of them, together with the scale factor, as a function of the other. The typical way to close the system is to equip Eq. (3.20) with an equation of state relating the pressure and the energy density $p = p(\rho)$.[31]

Notice that only the first equation of the system (3.20) is dynamical, since it is second order in the time derivatives; the second provides a consistency check for the initial conditions and the solution at every time.

There is a perfect similarity between the flat ($k = 0$) FLRW equations with zero pressure and the equations (2.9) and (2.12), found in the Newtonian framework for the dynamics of a homogeneous matter distribution. However, letting aside the inclusion of spatial curvature and relativistic matter, in that context it was necessary to overcome the divergences in the gravitational field by pretending that the potential is trivial inside a spherical cavity. Instead, in the context of relativistic cosmology this result naturally follows from the Birkhoff theorem (Weinberg 1972), no matter the extension of the matter distribution. Also, although the role of the scale factor $a(t)$ is conceptually the same, in the derivation conducted in the Newtonian context that function could only regulate the small distance evolution of the matter distribution. Instead, here any spacelike distance in the flat Friedman universe is regulated by

$$d(t) = a(t)r = \frac{a(t)}{a(t_0)} d(t_0), \qquad (3.21)$$

which is the analogous of Eq. (2.7).

---

[29] Another convention is, as usual, to attach to the radial coordinate the dimensions of a length and take the scale factor dimensionless. In this case the parameter $k$ can take any real value, since it is proportional to the Gaussian curvature of the constant $t$ surfaces.

[30] It should be mentioned that analogous equations can be obtained within the geometric formulation of Newtonian gravity discussed in the previous section, that is, Newton-Cartan theory (Cartan 1923,1924). This is especially useful when the Newtonian limit can give precious insights, as it happens for the cosmological perturbations after recombination (Ruede and Straumann 1997).

[31] For ordinary nonrelativistic matter (planets, stars, dust etc.) the pressure identically vanishes $p_{NR} = 0$, while for relativistic particles (e.g. photons and neutrinos) $p_R = \frac{1}{3}\rho_R$.

**3.4 Einstein closed and static universe**

The focus of this subsection is Einstein's solution for a closed and static universe, which historically led to the introduction of the cosmological constant. To this aim, I need to come back to his cosmological considerations (Einstein 1917).

As said, the mainstream assumption at the time was that the universe is static and spatially infinite, with a matter distribution appearing spherically symmetric when seen from our location. The debate hovered on whether matter is localized or extends indefinitely.

In either case, Einstein was confronted with the practical problem of laying down appropriate boundary conditions for the gravitational field at infinity. In his thinking, these not only had to be compatible with the astronomical observations but should also be compliant with the principle of Mach on the relativity of inertia (Mach 1893). The latter informed how the inertia of a body would not be an absolute and ontological property; rather, it would be completely determined by the matter distribution in the universe. This automatically excluded the most natural possibility of an asymptotically Minkowskian space, since at infinity it would define inertial frames independently from the matter distribution.

All this struggle convinced Einstein to drop altogether the idea of a universe extending indefinitely over space and consider, instead, a spatially compact universe with no boundaries, uniformly filled with nonrelativistic and static matter. In the cosmological model developed above (Eqs. (3.19) and (3.20)) this would correspond to the case $k = 1$ with constant scale factor and zero pressure; however, it is immediate to verify that such a model cannot be a solution of the Friedman equations.

It is for this reason that Einstein introduced a cosmological term in his field equations,

$$R_{\mu\nu} - \frac{1}{2} R g_{\mu\nu} + \Lambda g_{\mu\nu} = \frac{8\pi G}{c^4} T_{\mu\nu}, \qquad (3.22)$$

where the parameter $\Lambda$ would later be called "cosmological constant".[32]

Defining $\rho_\Lambda = \frac{\Lambda}{8\pi G}$, in the FLRW equations the new term has the same effect of a perfect fluid with equation of state $p_\Lambda = -\rho_\Lambda$

$$\begin{cases} \frac{\ddot{a}(t)}{a(t)} = -\frac{4\pi G}{3}[\rho_m(t) + 3p_m(t)] + \frac{\Lambda}{3}, \\ \left(\frac{\dot{a}(t)}{a(t)}\right)^2 + \frac{k}{a^2(t)} = \frac{8\pi G}{3} \rho_m(t) + \frac{\Lambda}{3}. \end{cases} \qquad (3.23)$$

---

[32] In the paper, he referred to the new term as a "slight modification" (Einstein 1917) of the field equations; he could not know that it would have become one of the greatest problems of contemporary physics. Also, notice that the cosmological term is compatible with the principles of general relativity and the covariant conservation of the energy momentum tensor.

The modified system now admits Einstein's universe as a solution, with a constant scale factor[33] $a = \frac{1}{\sqrt{\Lambda}}$.

A central point of the next subsection will be Hubble's discovery of the expansion of the universe, which definitely discarded static solutions. Nonetheless, one can realize that Einstein's universe cannot be a stable solution purely on a theoretical ground (Weinberg 2008). Consider small perturbations of the matter density, $\rho_m = \frac{\Lambda}{4\pi G} + \delta\rho_m(t)$, and of the scale factor, $a = \frac{1}{\sqrt{\Lambda}} + \delta a(t)$. To first order in the perturbations, the system (3.23) reads

$$\begin{cases} \ddot{\delta a} = -\frac{4\pi G}{3} \frac{1}{\sqrt{\Lambda}} \delta\rho_m, \\ \Lambda \delta a = -\frac{4\pi G}{3} \frac{1}{\sqrt{\Lambda}} \delta\rho_m, \end{cases} \quad (3.24)$$

and yields $\ddot{\delta a} = \Lambda \delta a$, $\ddot{\delta\rho}_m = \Lambda \delta\rho_m$. Therefore, as soon as a perturbation in the energy density appears, it starts growing exponentially (exponential instability) and eventually breaks the solution.

### 3.5 Expansion of the universe

Equation (3.19) suggests that the dynamics of the universe can be deduced by looking at the proper distance of astrophysical objects from Earth, since the latter depends on the scale factor

$$d(r, t) = a(t) \int_0^r \frac{dr'}{\sqrt{1-kr'^2}}. \quad (3.25)$$

In particular, a galaxy should appear receding from us if the universe is expanding, whereas it should appear approaching if the universe is contracting.

If we could lay down rulers from Earth up to a different galaxy (that is, directly measure the proper distance), the scale factor could be easily deduced. Instead, true observables need to be found which relate the distance of an object with the dynamics of the universe.

Consider a luminous galaxy emitting photons at a definite frequency $\nu_{em}$.[34] As the spatial distances change in the travel towards the Earth, the wavelength of the radiation is expected to change accordingly because the number of oscillations is independent from the gravitational field. Hence, the observed frequency $\nu_0$ in general is different. The quantity conventionally introduced to assess the frequency shift is (Wald 1984)

---

[33] Notice how this is compliant with the perfect cosmological principle in its strongest interpretation, in which all the physical quantities are constant. Instead, the steady state cosmology of Bondi, Gold and Hoyle was based on a weaker interpretation, in which only the observables are assumed constant (Weinberg 1972).

[34] Sources for which the frequency of emission is known (because the nuclear processes are) are called standard candles (Weinberg 2008).

$$z \equiv \frac{\nu_{em} - \nu_0}{\nu_0} = \frac{a(t_0)}{a(t_{em})} - 1, \tag{3.26}$$

and can be observed. It is also conventional for small shifts ($|z| \ll 1$) to assign the apparent velocity $v_z = cz$ to the source, as if the change in the frequency were due to Doppler effect in flat spacetime (Weinberg 1972).

As for the distance, the overwhelmingly most used indicator in cosmology is the luminosity distance (Weinberg 2008). This is deduced from the knowledge of the absolute luminosity of the source $L$, that is the energy per unit time emitted as radiation; and from the measure of the apparent luminosity $l$, which is the energy per unit time and surface detected by the observer

$$d_L = \sqrt{\frac{L}{4\pi l}}. \tag{3.27}$$

It turns out (Weinberg 2008) that for small frequency shifts in the emitted radiation the luminosity distance can be written as a power series in $z$

$$d_L(z) = \frac{1}{H(t_0)} \left\{ z + \frac{1}{2}[1 - q(t_0)]z^2 \right\} + o(z^3). \tag{3.28}$$

In this expression $H(t) \equiv \frac{\dot{a}(t)}{a(t)}$ and $q(t) \equiv -\frac{\ddot{a}(t)}{H(t)^2 a(t)}$ are the Hubble and the acceleration parameters, both evaluated at observation time, and can be estimated by fitting the measurements of luminosity distance and frequency shift.

Retaining only the first order in $z$ one gets

$$d_L(z) \simeq \frac{1}{H(t_0)} z, \tag{3.29}$$

which is the renowned Hubble law.

In 1929, slightly more than a decade after Einstein's theorization of a static universe, Hubble claimed the discovery of a "roughly linear relation" (Hubble 1929) between the observed positive shifts (i.e. redshifts, $z > 0$) and the luminosity distances of nearby galaxies, with a Hubble parameter estimated about $H(t_{obs})|_{Hubble} \simeq 500 \frac{km}{sMpc}$. The extrapolation was based on a sample of 24 galaxies, all within the Virgo cluster ($z \leq 0.01$), and would have appeared extremely problematic today. The reason is that, other than the apparent velocity induced by the large scales dynamics, galaxies possess "peculiar" velocities which are determined by the local gravitational field, independently from the cosmological evolution. Such components hover around $\sim 10^3 \frac{km}{s}$; therefore, to have a dominant cosmological contribution one would need to look at shifts greater than 0.01. This is the reason why the 1929 measurements employed by Hubble were spread out in the $d_L$-$z$ plane and could not support a linear relation with great statistical significance. But luckily enough this conclusion turned out to be correct: by the early 1930's Hubble had pushed the observations up to galaxies in the Coma cluster ($z \simeq 0.02$), and a linear relation with a positive Hubble parameter was now evident. From that point onward there have been no more doubts: the universe is expanding, and the determination of $H(t_0)$ has become a typical challenge in high precision astronomy. At the time of writing, one of the latest measurements has been performed by Riess et al. in 2022 (Riess et al 2022), using observations of Supernovae Ia from the Hubble Space Telescope (HST) $(H(t_0)|_{HST} = (73.30 \pm 1.04) \frac{km}{sMpc})$.[35]

The first practical consequence of the expansion of the universe is that $z$ is indeed a cosmological redshift; hence, distant sources appear receding from us. At a more conceptual level, this discovery really marked the end of the scientific revolution (Kuhn 1962) initiated by Einstein three decades before, which led from an immutable universe governed by Newtonian forces to an evolving universe governed by the interplay between energy and spacetime geometry.

But since the universe is not static, what about the cosmological constant? Should it just be thrown away?

The response of Einstein himself would have certainly been in the positive as already in 1923, in a letter to Hermann Weyl he was writing: "If there is not any quasi-static universe after all, then get rid of the cosmological term" (Einstein 1923). However, the genie had been left out of the bottle and could no longer easily be forced back in (Zeldovich 1968).

### 3.6 Accelerated expansion

To see why the cosmological constant could not be just washed away, let me fast forward to the end of the century when the discovery of the accelerated expansion took place.

From Eq. (3.28) the logical next step would be to push the observations to higher redshifts and fit the acceleration parameter $q_0$. More than that, one could carry the expansion of the luminosity distance up to higher powers of $z$ and backtrack $a(t)$ by fitting its higher derivatives. However, this would require a lot of high precision measurements at intermediate redshifts (greater than 0.1 but small enough not to break the small $z$ approximation), hardly available in astronomy.

What is done instead is to use (not that precise) measurements in a wide range of redshift up to unity and fit the exact form of the function $d_L(z)$. This has an integral expression (see Weinberg 2008) depending on $H(t_0)$ and the four parameters

$$\Omega_k \equiv -\frac{k}{a^2 H^2}, \quad \Omega_{NR} \equiv \frac{8\pi G \rho_{NR}}{3H^2}, \quad \Omega_R \equiv \frac{8\pi G \rho_R}{3H^2}, \quad \Omega_\Lambda \equiv \frac{8\pi G \rho_\Lambda}{3H^2}, \qquad (3.30)$$

linked to the energy densities. The second equation of the system (3.23) requires that the $\Omega$'s sum up to unity at every time; therefore, those parameters inform how the total energy density in the universe is shared among the various form of matter and the curvature of space.

---

[35] Another recent determination is by (Balkenhol et al), using measurements of the CMB power spectrum from the Plank satellite. The result is $H(t_0)|_{Plank} = (68.3 \pm 1.5) \frac{km}{sMpc}$, more than $5\sigma$ distant from the HST observations. More generally, $H(t_0)$ measurements from supernovae Ia seem not compatible with the CMB determinations. Understanding the reason for this tension is an open problem in cosmology (Perivolaropoulos and Skar 2022).

Between 1998 and 1999, the Supernova Cosmology Project (Perlmutter et al 1999) and the High-z Supernova Search Team (Riess et al 1998) analysed the $d_L$-$z$ relation of, respectively, 42 and 16 type Ia supernovae, within a redshift range closely reaching unity. Comparing with the theoretical prediction both teams established

$$\Omega_k = 0, \ \Omega_{NR} \simeq 0.28, \ \Omega_R = 0, \ \Omega_\Lambda = 0.72 \qquad (3.31)$$

at more than 99% confidence level. This is a discovery of enormous importance, because it tells that the energy density in the universe is currently dominated by the cosmological constant, followed by nonrelativistic matter with the same order of magnitude.[36] From the first of the FLRW equations, this immediately implies that the expansion of the universe is accelerating.

Also, from the definition of $\Omega_\Lambda$ the energy density of the cosmological constant is assessed to the value

$$\rho_\Lambda \simeq 10^{-47} \text{GeV}^4, \qquad (3.32)$$

a result which will be central in the discussion of the next section.

It is remarkable how the cosmological constant, initially introduced by Einstein to fill a static universe, turns out to be the chief motor of its expansion. The question urges: what is the physical nature of $\Lambda$?

It is unsettling that at the end of the path outlined up to now the answer to this question still dramatically lacks. The next section addresses this problem, together with its deep connection with the quantum theory of fields.

## 4 The cosmological constant problem

This last section is devoted to the cosmological constant problem. The first part discusses the coupling of the quantum fields with the Einstein equations. The second part addresses the computation of the vacuum energy of the fields as foreseen by the standard model of particle physics, showing the great discrepancy with the value deduced from the cosmological observations. In the last subsection some key conceptual issues of the problem are discussed. The resources are (Mukhanov and Winitzki 2007), (Martin 2012), (Weinberg 1995, 1996, 2013) and (Weinberg 1989).

### 4.1 Semiclassical gravity and vacuum energy

Besides the path leading to general relativity, the first decades of the 20th century hosted another revolutionary process in physics: the development of quantum mechanics. This took off from the works of Plank (Planck 1901), Einstein (Einstein 1905b) and Bohr (Bohr 1913), even though a more systematic formulation of the theory had to wait the works of Heisenberg (Heisenberg 1925), Schrodinger (Schrodinger 1926) and others. In this context, the first quantized field appeared in 1926[37] and it was the electromagnetic field. Immediately after, the same quantization method (second quantization) was applied to spacetime functions supposedly representing the classical wave functions for other types of particles (e.g. the electron). However, the true birth of the quantum fields[38] only happened when, first, it was realized that those functions cannot be understood as probability densities. Those are entirely separate entities (classical fields), which after quantization describe the creation and annihilation of multiple particles and antiparticles. And second, the field equations were formulated in a Lorentz covariant fashion (Weinberg 1995).

---

[36] The resulting model takes the name of $\Lambda$-CDM; the second part standing for cold dark matter, the other mystery of contemporary cosmology (Bertone and Hooper 2018). Moreover, understanding why the cosmological constant and the nonrelativistic matter contributions are of the same order is known as the coincidence problem (Amendola and Tsujikawa 2010).

Today, matter particles and their interactions are described by the standard model of particle physics (Cottingham and Greenwood 2007). In this framework, three of the four fundamental forces (weak, electromagnetic and strong) are described in terms of exchanges of mediating particles (e.g. the photon for the electromagnetic force), while gravity is kept "classical" and described by general relativity.

Here, it suffices to say that a field is a function[39] of the spacetime coordinates which defines, at any point, an operator acting on the particle states of the system. The latter describe the types of particles that are present, together with their energy and momentum. In particular, the vacuum state $|0\rangle$ indicates the absence of any propagating particle and it is the minimum energy state of the system.

Since any form of energy gravitates, on the right-hand side of the Einstein equations (3.22) one should acknowledge the contribution of the quantum fields. This is a nontrivial step because the energy momentum tensor itself is an operator, which must be embedded in a classical equation. It turns out that (Wald 1994) if the quantum fluctuations are contained, the coupling can be realized by taking the vacuum expectation value[40] $\langle \hat{T}_{\mu\nu} \rangle \equiv \langle 0 | \hat{T}_{\mu\nu} | 0 \rangle$

$$R_{\mu\nu} - \frac{1}{2} R g_{\mu\nu} + \Lambda g_{\mu\nu} = \frac{8\pi G}{c^4} \left( T_{\mu\nu}^{cl} + \langle \hat{T}_{\mu\nu} \rangle \right). \tag{4.1}$$

In these semiclassical Einstein equations the term $T_{\mu\nu}^{cl}$ stands for the contribution of sources which can be kept classical.

From the field theoretic viewpoint, the difference between the two sides of Eq. (4.1) is revelatory: on the right-hand side all the machinery of the quantum theory is needed to describe the matter fields sourcing gravity, while on the left-hand side the gravitational field itself is kept classical. At the time of writing, whether this is a fundamental difference it cannot be answered definitively. In fact, understanding the nature of the gravitational field within the framework of the quantum theory is probably the greatest challenge in current theoretical physics (Giulini, Kiefer and Lämmerzahl 2003), (Stachel

---

[37] See (Weinberg 1995) for a historical introduction to the quantum theory of fields.

[38] See (Peskin and Schroeder 2018) and (Maggiore 2005) for an introduction to the quantum theory of fields; and (Weinberg 1995, 1996), (Zinn-Justin 1996) for more advanced expositions.

[39] A more rigorous definition would tell that a field is a tempered distribution of the spacetime coordinates (Weinberg 1995), which is well defined only as a linear functional of differentiable and rapidly decreasing functions.

[40] The hat indicates that the quantity is an operator, as usual in quantum mechanics. See (Mukhanov and Winitzki 2007) for a derivation.

1999): loop quantum gravity[41] (Thiemann 2007), asymptotic safety[42] (Percacci 2017) and string theory[43] (Green, Schwarz and Witten 1987) are just three of the most authoritative approaches deserving a mention. The operative posture on the matter adopted here is that, at the energies at play for the cosmological constant problem, the possible hidden physics at the Planck scale[44] ($E_p \simeq 10^{19} GeV$) may not be crucial and semiclassical gravity may be a good effective theory.

Coming back to Eq. (4.1), notice that the effect of the quantum fluctuations should be intended in a perturbative sense, because the definition of the zero-particle state depends on the background metric (Mukhanov and Winitzki 2007). At the operational level, one can find a solution $g_0$ of the equations without the quantum contributions and use the latter to define the zero-particle state $|0\rangle_{g_0}$. Then one considers Eq. (4.1) and finds the correction to $g_0$, $g_1 = g_0 + \delta g_0$, induced by the fluctuations.

The procedure can be iterated, defining new vacua and finding new corrections to the metric, but it may not converge if the fluctuations start to grow.

Going back to $\langle \hat{T}_{\mu\nu} \rangle$, the local Lorentz invariance of the vacuum state (Zel'dovich and Krasinski 1968) forces the expectation value to take the simple form

$$\langle \hat{T}_{\mu\nu} \rangle = -\rho_{vac} g_{\mu\nu}, \qquad (4.2)$$

where $\rho_{vac}$ is the constant vacuum energy density. This is remarkable because it shows that the zero point energies of the fields act exactly as a cosmological term on the spacetime geometry. Therefore, in Eq. (4.1) it is natural to gather the contributions in an effective cosmological constant

$$\Lambda_{eff} = \Lambda + \frac{8\pi G}{c^4} \rho_{vac}. \qquad (4.3)$$

The reason for not dropping $\Lambda$ will be clear in a moment, once the value of $\rho_{vac}$ has been computed.

**4.2 The cosmological constant problem**

From an energetic point of view a field can be seen as a collection of quantum harmonic oscillators (Konishi and Paffuti 2009), one at each spatial point and for every value of the spatial momentum $\vec{p}$.[45]

---

[41] This is based on a direct application of Dirac's quantization program to the Hamiltonian formulation of general relativity (Thiemann 2007).

[42] Such approach is based on the idea that, although the quantum theory of a propagating graviton is nonrenormalizable, physical information can nonetheless be extracted near a possible ultraviolet fixed point in the variation of the coupling constants with the energy scale (Percacci 2017). Together with loop quantum gravity, asymptotic safety is among the leading nonperturbative approaches to the quantization of gravity.

[43] In which the quantization of the gravitational field nests within the broader program of unification of all fundamental interactions (Green, Schwarz and Witten 1987).

[44] This is the energy scale at which the possible quantum effects of the gravitational field are expected to be nonnegligible (Giulini, Kiefer and Lämmerzahl 2003).

[45] If the field has spin, there is a collection of harmonic oscillators for every spin degree of freedom.

Therefore, in the simplest case of a scalar field of mass $m$ the vacuum energy reads ($|\vec{p}| = p$)

$$\rho_{vac} = \int \frac{d^3\vec{p}}{(2\pi)^3} \frac{1}{2} \omega(|\vec{p}|) = \frac{1}{(2\pi)^2} \int_0^{+\infty} dp \; p^2 \sqrt{p^2 + m^2}. \qquad (4.4)$$

The integral is divergent as $\sim p^4$, leading to an infinite vacuum energy. In the absence of gravity this would not be a problem, because the experiments involving the matter fields can only probe differences in energy between particle states. But when the interaction with the gravitational field is considered, things become drastically different: the vacuum energy now gravitates and should have observable effects. This is a first example of the problem of divergences in quantum field theory (Weinberg 1995).

Equation (4.4) can manifest, either, that the theory may not be able to describe particles with infinitely large momenta, or that $\rho_{vac}$ is not an entirely physical quantity. It is crucial in this respect that the problem only arises when the gravitational interaction is "switched on".

Consider the electromagnetism in a medium as an example. One can write formulas using the speed of light in vacuum $c$; however, at some point it becomes convenient to define the speed of light in the medium $c_n = \frac{c}{n}$, with $n$ the refraction index, and reparametrize the theory using the latter. Since experimentally we do have access to light propagating in vacuum and in a medium, both $c$ and $c_n$ are physical and finite. But imagine we could only access electromagnetic phenomena in a medium, say the atmosphere of the Earth. One could still formulate the theory in vacuum and then introduce the interaction with the atmosphere; however, since $c$ would not be observable it would not need to be finite. The definition of the physical quantity $c_n$ in terms of $c$ would still be possible, but the coefficient $n$ could be infinite. Therefore, if the observables of the interacting theory are computed using $c$, one should expect divergences due to the "poor choice" of parametrization and reabsorb them in the definition of $c_n$.

This is a fictitious example of renormalization (Weinberg 1995, 1996), but it is exactly what happens in quantum field theory when the interactions among the fields are considered. The structure of the divergences is usually very complicated, with many unphysical parameters (bare parameters) entering the theory; therefore, in the typical cases one makes the observables finite with the introduction of regulators (e.g. a momentum cutoff in (4.4)). One then performs a convenient, collective reparametrization of the bare quantities in terms of the physical ones (renormalized parameters), in order to reabsorb the divergences, and finally removes the regulators (e.g. taking the limit of infinite momentum cutoff).

In the case of the gravitating vacuum energy the quantity in Eq. (4.4) is playing a role analogous to the bare parameter $c$, because it pertains to a context in which there is no gravity. Using a dimensional regulator (Weinberg 1995), the finite bare vacuum energy turns out (Martin 2012)

$$\rho_{vac}^B = -\frac{m^4}{64\pi^2} \left[ \frac{2}{\epsilon} + \frac{3}{2} - \gamma - \ln\left(\frac{m^2}{4\pi\mu^2}\right) \right] + o(\epsilon). \qquad (4.5)$$

In this expression $\epsilon = 4 - D$, with $D$ the number of spacetime dimensions; $\gamma$ is the Euler-Mascheroni constant; and $\mu$ is an arbitrary mass scale inserted for dimensional reasons. Removing the regulator corresponds to the limit $D \to 4$, in which case Eq. (4.4) is recovered.

The physical (or renormalized) vacuum energy is hidden in the finite part of $\rho_{vac}^B$, that is

$$\rho_{vac}^B = \rho_{vac}^{ren} + \delta\rho_{vac}^B, \tag{4.6}$$

where $\delta\rho_{vac}^B$ incorporates the divergence. Since only $\rho_{vac}^{ren}$ can enter the Einstein equations, a comparison with Eq. (4.3) reveals that $\Lambda_{eff} = \frac{8\pi G}{c^4}\rho_{vac}^{ren}$, while

$\Lambda = \frac{8\pi G}{c^4}\delta\rho_{vac}^B$. This observation is very powerful and implies a further fundamental change in the meaning of the original cosmological constant, acting now as the divergent counter-term in the renormalized vacuum energy.

But how exactly should one identify the physical contribution within the finite part of Eq. (4.5)?

This is a very sensible point lying at the heart of renormalization. The answer is that there is not a unique way to subtract the divergences with only the theory at hand. One can choose a rule and apply it consistently to every observable, but at the end of the day the physical quantities will always be defined up to the addition of finite contributions (i.e. finite reparametrizations). A one-to-one correspondence with nature can only be established stepping out of theory and imposing that, by definition, the given renormalized quantity corresponds to the outcome of an experiment.[46] This fixes once and for all the finite contributions.

Such observation somewhat clarifies why the finite part of $\rho_{vac}^B$ depends upon the energy scale $\mu$. The point is that, of all the conditions influencing the experiment, within the renormalization what really matters is the typical energy scale at which the experiment is performed. Think at the weak interaction of two electrons via the exchange of a $Z_0$ boson. If the energy scale of interest is much less than $M_{Z_0} \simeq 91$ GeV, the process simply does not happen and one can consider an effective theory in which the electrons only interact electromagnetically. But as argued, renormalization is determined by the presence and the type of interactions; hence, the theory "naturally asks" for the introduction of an energy scale in the physical parameters. The latter is $\mu$ in Eq. (4.5).

Coming back to the question above, the most natural choice to identify $\rho_{vac}^{ren}$ is to subtract the pole in Eq. (4.5) together with the finite terms which do not depend on $\mu$; so that

$$\rho_{vac}^{ren} = \frac{m^4}{64\pi^2}\ln\left(\frac{m^2}{4\pi\mu^2}\right) + \text{(finite reparam.)} \tag{4.7}$$

This is the contribution of a single scalar field, but the generalization to multiple free fields is straightforward (Martin 2012)

---

[46] The comparison with an experimental value which fixes the finite contributions is called a renormalization condition.

$$\rho_{vac}^{ren} = \sum_i n_i \frac{m_i^4}{64\pi^2} \ln\left(\frac{m_i^2}{4\pi\mu^2}\right) + \text{(finite reparam.)}. \tag{4.8}$$

Here the sum is extended to all the massive fields while $n_i$ counts their internal degrees of freedom, with a positive sign for bosons and a negative one for fermions. To get a realistic prediction, one now needs to specify the fields and the energy scale corresponding to an experiment.

As for the first, according to the standard model there exist three leptons (electron, muon and tauon, $n_l = -4$) with $m_e \simeq 0.5$ MeV, $m_\tau \simeq 1.8$ MeV and $m_\mu \simeq 105$ MeV; the six quarks ($n_q = -4$) with $m_t \simeq 171.2$ GeV, $m_b \simeq 4.2$ GeV, $m_d \simeq 0.5$ GeV, $m_u \simeq 0.2$ GeV, $m_s \simeq 0.1$ GeV and $m_c \simeq 1.3$ GeV; the neutrinos, whose possible mass is of the order of the electronvolt (Cottingham and Greenwood 2007); the three bosons mediating the weak interaction ($W^\pm$ and $Z_0$, $n_{W,Z} = 3$), with $m_W^\pm \simeq 80$ GeV and $m_Z \simeq 91$ GeV; the mediators of the electromagnetic and the strong interaction (gluons), which are massless; and the Higgs boson ($n_H = 1$), with $m_H \simeq 125$ GeV.

As for the second, from the previous section it is clear that the experimental data is provided by the observations of high redshift supernovae, Eq. (3.32); however, in this case the energy scale is not "controlled" as for scattering experiments in accelerators. Taking into account the energy of the photons coming from the supernovae, whose wavelength is about $\lambda \sim 500$ nm, and the energy scale induced by the expansion of the universe ($H_0 \simeq 70 \frac{\text{km}}{\text{s}} \frac{1}{\text{Mpc}}$), a plausible estimate is (Martin 2012)

$$\mu^* = \sqrt{E_\gamma E_{grav}} \simeq 3 \times 10^{-25} \text{GeV}. \tag{4.9}$$

Putting together the last observations, the theoretical prediction for the vacuum energy in the cosmological setting is

$$\rho_{vac}^{ren}(\mu^*) \simeq -2 \times 10^9 GeV^4 + \text{(finite reparam.)}, \tag{4.10}$$

which should be compared with the experimental result

$$\rho_\Lambda \simeq 10^{-47} \text{GeV}^4. \tag{4.11}$$

Apart from the sign, the two values differ for 56 orders of magnitude and would require an incredible degree of fine-tuning in the choice of the finite reparametrization.
This is the essence of the cosmological constant problem.

**4.2 Discussion**

There are several things to discuss about the results of the last subsection.

First, let me remark that the calculation of the vacuum energy in quantum field theory is only one aspect of the problem. Another aspect is connected with the phase transitions[47] occurring in the early universe, as the temperature cools down (Peter and Uzan 2009). These cause a significant shift in the vacuum energy of the fields; hence, the choice of the finite contribution in Eq. (4.8) strongly depends on whether the renormalization condition is fixed before or after the phase transition.

A second point concerns the issues of the working hypotheses behind the expression (4.8). In particular, the computation has been performed for a collection of free fields in a flat spacetime, whereas one should consider the interacting fields of the standard model in a curved background. The reasons for this choice are sensible even if not completely satisfactory. Regarding the interactions, it has been argued how they affect the renormalization of the parameters of the theory; hence, in Eq. (4.8) the renormalized masses of the fields should appear, evaluated at the energy scale of the cosmological observations. This would affect the value of the vacuum energy but it is not expected to provide a radical change.[48] As for the coupling with a curved spacetime, it should be mentioned that a consistent formulation of a quantum field theory in a curved background is in and of itself problematic (Wald 1994). Apart from that, since the main contributions to the vacuum energy come from the high momenta of the fields, one could argue that at those scales the curvature of a regular spacetime can be neglected and the computation on Minkowski space is justified by the equivalence principle. However, the problem with this point of view is that it is assuming the existence of a nondynamical background.

In the reasoning leading to Eq. (4.1) a crucial point is that the quantum fields should back react on the metric in a perturbative way; therefore, a large concentration of their energy not only would end up shaping the background solution but would also break the perturbative approach.[49] This is a major point which should need much more study and, in fact, it is even more fundamental than the cosmological constant problem, since it roots into the very definition of a quantum field theory on a dynamical background.

Let me finally come back to the fine-tuning problem stated at the end of the last subsection. One may wonder why, after all the computation, one should care about the fact that an unphysical parameter must be chosen with such great accuracy: after all, if it is unphysical, can it not just be fixed to whatever value it is needed? The point is that, because of the consequentiality through which we understand the laws of nature, it is unlikely that a physical quantity be explained by the accurate choice of the parameters of our theories. There is no other way to put it. Also, history provides countless examples that, when fine tuning is heavily needed in a theory, something is going wrong.[50]

---

[47] The electroweak phase transition is a famous example, in which the unified electroweak interaction splits into the electromagnetic and the weak interactions (Weinberg 1996)

[48] For the electron mass, the difference between the value $m_e$ inserted in Eq. (4.8) and the (one loop) renormalized mass evaluated at $\mu^*$ is of the order of $m_e$ (Schwartz 2014).

[49] This observation came from S. L. Cacciatori in one of our many private discussions.

[50] Take the ether theory for instance. At some point, the dragging models became so artificial that someone ended up exploring a different and more natural direction, dropping in the process a hypothesis which had been in place for centuries.

An equivalent, possibly more rigorous way to address the problem is by acknowledging that the cosmological constant is very small compared to the vacuum energy of the quantum fields. In a sense proposed by t' Hooft in the 1980s ('t Hooft 1980), a parameter is allowed to be very small at some energy scale only if there exists an underlying symmetry which is minimally broken by its nonzero value.[51] But since there are no convincing evidences of a symmetry underlying the vacuum energy, the value of the cosmological constant deduced from the observations appears unnatural.

There is another fold of the problem which is worth to mention. Going back to the discussion on the renormalization, it is important to clarify that a nonrenormalizable theory is not automatically unphysical or ill defined; rather, it is a theory with an explicit cutoff energy scale, which is not able to provide predictions at higher energies. On the other hand, a renormalizable theory can make predictions at all energies. Does it mean that we can trust those predictions all the way up to infinitely great energies? Of course not! We typically expect new physics to lie down there, possibly in the form of new massive fields getting excited.[52] One could reverse the argument and ask whether the low energy predictions can be trusted: if there is new physics, how can we be sure that it does not affect the predictions at lower energy? The answer is that a renormalizable theory should not be sensitive to the ultraviolet regime. In other words, the theory must behave in such a way that its predictions do not depend on the details at infinitely small length scales. Looking at Eq. (4.8), it is clear from the dependence upon the mass of the particles that if new massive fields get excited, the renormalized vacuum energy can change significantly. And because the observed value is so small the finite terms should be tuned anew to a totally different value. This is a manifestation of great sensibility to the ultraviolet.

## 5 Perspectives and conclusions

Up to this moment, the many facets of the cosmological constant problem have been outlined; it is now time to address the attempts for its solution.

In line with t'Hooft argument, a first idea involves the formulation of a minimally broken hidden symmetry to explain the smallness of the cosmological constant. The most natural assumption follows from the fact that in Eq. (4.8) the contributions from bosons and fermions have opposite sign; therefore, an exact symmetry in nature between the two species would lead to the cancellation of the vacuum energy. The idea of a symmetry between bosonic and fermionic particles (or supersymmetry) appeared for the first time in the early 1970, while the observation that it could naturally regulate the vacuum energy came from Zumino thereafter. In more recent times, the failing of the large hadron collider to find any supersymmetric partner has damped the enthusiasm towards this possibility; nonetheless, the study of supersymmetric theories is still object of active research efforts (Weinberg 2013).

---

[51] The typical example is the mass of the electron, which breaks the chiral symmetry (Weinberg 1996)

[52] For instance, some new very massive fields unifying the standard model (Cottingham and Greenwood 2007).

A second possibility consists of those attempts acting at the level of the semiclassical equation (4.1), with the fundamental idea of making gravity strongly less sensitive to the vacuum energy of the fields (Weinberg 1989). The typical mechanisms to do so are the so-called adjusting mechanisms, in which a field drives the vacuum energy to zero in its evolution towards a stable configuration. In this respect, an important result is provided by a no-go theorem due to Weinberg (Weinberg 1989). It states that, under some quite general assumptions about the field theory of matter and gravity, an adjusting mechanism is not possible without fine tuning the parameters of the theory. This result is very powerful, because it indicates the set of conditions which must be weakened to avoid fine tuning.

Another popular route which deserves a mention is to modify or extend general relativity[53] (see (Capozziello and Faraoni 2010), (Amendola and Tsujikawa 2010)) in order to reproduce the acceleration of the universe purely as a gravitational effect.[54] This does not address directly the cosmological constant problem, but it does decouple the latter from the explanation of the accelerated expansion. In a successful scenario one should then understand what are the effects of such a great value of the vacuum energy on the cosmological expansion.

In some sense, the three research lines outlined above do not challenge the basic assumptions of modern cosmology or the fundamental particle physics framework;[55] however, very interesting possibilities lie beyond such foundations.

For example, many compactifications of higher dimensional string theories seem to admit a stable anti-de Sitter (AdS) vacuum (Kachru et al. 2003); that is, a FLRW universe with a negative cosmological constant. But since the experimental evidence of a small and positive cosmological constant, many efforts have been devoted by the string community to construct sufficiently meta-stable de Sitter (dS) vacua[56] (Kachru et al. 2003). To date, whether such dS states actually belong to the string theory landscape is still matter of debate (Obied et al. 2018).

On this note, it should be mentioned that a transition from an early AdS state to the present dS phase at a proper red shift would alleviate both the $H_0$ and the $S_8$ tensions[57] of the Λ-CDM model (Akarsu et al. 2021).[58] Models envisioning such a transition involve a dynamical cosmological "constant" $\Lambda_s$ (or dynamical dark energy), acting as an order parameter in a phase transition, and switching sign at a proper redshift.[59] Notice that such a mechanism would be compliant with t'Hooft argument on the necessity of a broken hidden symmetry to explain why the cosmological constant is so small.

---

[53] As discussed in the former section, general relativity is deeply rooted in three physical principles: that of general covariance, the relativity and the equivalence principles; the latter being confirmed experimentally with great accuracy in the Einstein's formulation (Will 2014). To date, no gravitational theory other than general relativity is known to satisfy the equivalence principle in its strongest formulation (Di Casola, Liberati and Sonego 2015).

[54] That is, without the need of the cosmological constant in the matter sector.

[55] Even in modified gravity theories the cosmological principle is usually assumed and, at least for the theories obeying the Einstein formulation of the equivalence principle (Di Casola, Liberati and Sonego 2015), the coupling with the matter fields is realized through the metric tensor (Capozziello and Faraoni 2010).

[56] That is, a flat FLRW universe with a positive cosmological constant and a lifetime greater than the cosmological timescale.

[57] The $S_8$ tension refers to the discrepancy between the high and low redshift determinations of the strength of the matter clustering, see (Abdalla et al. 2022).

Lastly, a very interesting avenue opens as soon as the cosmological principle is challenged. From Sec. 3 it is evident that isotropy and (especially) homogeneity are effective spatial symmetries, emerging on average over sufficiently large distances. However, the somewhat tacit assumptions carried along by the standard model are, on the one hand, that the small scales violations of those symmetries have negligible impact on the large scale evolution of spacetime; and on the other, that the reference frame of typical observers like ourselves coincides with the coarse grained reference frame defined by the cosmological principle. That these two assumptions may not prove accurate is the idea motivating the inhomogeneous cosmological models (Bolejko and Korzynski 2017).

A first interesting approach was pioneered by Pietronero in the 1980's (Pietronero 1987) and consisted in modelling the observed hierarchical structure of galaxies and clusters with a fractal system.[60] The approximate gravitational field generated by such a fractal structure was proposed by Ribeiro some years later (Ribeiro 1992) in the form of a Lemaître-Tolman-Bondi (LTB) metric[61] (Stephani et al. 2003). In this view, the resulting global geometry of spacetime would be a "Swiss cheese" of LTB regions nested in an otherwise FLRW universe. As was recently shown by Pietronero and collaborators (Cosmai et al. 2018), in such geometry the supernovae Ia observations are explained without the need of dark energy,[62] so that the apparent acceleration of the cosmic expansion would be an artifact of the homogeneous cosmology.[63]

---

[58] According to the AdS distance conjecture (Lüst, Palti and Vafa 2019) this would be highly unlikely in the string theory framework. However, it was recently argued that such a transition could in fact be driven by Casimir-like forces among the fields inhabiting the bulk of a 5D spacetime (Anchordoqui, Antoniadis and Lüst 2024). Therefore, the feasibility of the AdS – dS transition in string theory is still an open question.

[59] Notice that, within the standard cosmological model developed in Sec. 3, a changing in the sign of the cosmological constant necessarily implies a change in the topology (that is, $k$ in Eq. (3.13)) of the constant time hypersurfaces, since AdS does not admit a flat slicing. In some models the sign change of $\Lambda$ is realized with constant $k$, by admitting a sign flip in the signature of the Lorentzian metric (that is, from mostly plus to mostly minus) (Alexandre, Gielen and Magueijo 2024). In these cases, the definition of the Gaussian curvature of the hypersurfaces changes accordingly: from positive to negative for the 3-sphere ($k = +1$) and vice versa for the hyperbolic 3-space ($k = -1$).

[60] In a sense, this could be considered a modern version of Charlier's hierarchical cosmology, see e.g. (Norton 1998).

[61] This was the first isotropic but non homogeneous exact solution of the Einstein equations; it represents a spherically symmetric dynamical universe with a definite centre. See also (Cacciatori, Marrani and Re 2020) for a generalized LTB solution modeling a fractal system.

[62] Strictly speaking, this is the case when the fractal dimension approaches the actual dimension of space (D = 3) and when the typical length is about ∼ 100 Mpc (Pietronero et al.).

[63] In fact, the expansion would be slowing down in this model (Cosmai et al. 2018).

31An analogous conclusion about the accelerated expansion is reached in the context of the "timescape cosmology" proposed by Wiltshire in 2007 (Wiltshire 2007). This is a solution to the Buchert's equations[64] (Buchert 2000) based on a concrete distinction between the reference frame of a volume averaging observer, that is an observer averaging over the whole observable space, and that of an observer living in a high density region (or wall), like ourselves. Specifically, because of the local gravitational field and the presence of cosmic voids, the clock of a wall observer[65] runs slower than a volume averaged clock; as a consequence, conditions can be realized in which a wall observer measures an apparent acceleration, while on average none is detected. A first data analysis which may favour timescape cosmology with respect to the standard Λ-CDM was conducted in 2024 (Lane et al. 2024).

Finally, the η-CDM model recently proposed by Lapi and collaborators deserves a mention (Lapi et al. 2023). In this approach, matter inhomogeneities are modelled by adding a proper stochastic noise term in the FLRW equations, thus inducing slight statistical differences in the evolution of different patches (∼ tens of Mpcs) of the universe. When the system is averaged over the whole ensemble of patches, the effective equations of the overall dynamics are obtained. The latter show that a noise-induced acceleration at late times can be realized without the need of dark energy.

All these approaches challenge in different ways the standard interpretation of the Copernican principle, which essentially envisions a complete equivalence of the observers at any spatial location. Specifically, both the fractal and the timescape cosmologies introduce a difference between material points, belonging to high density regions, and points within the voids. In Pietronero's theory an equivalence can be established only among points belonging to the fractal system, thus recovering a "weaker" interpretation of the principle also called "conditional Copernican (cosmological) principle" (Mandelbrot 1983). On the other hand, Wiltshire argues that a void observer would actually see the same geometrical features as a wall observer, although with different outcomes of specific measurements.[66] In the $\eta$-CDM model, points belonging to different patches undergo slightly different evolutions but the Copernican principle is essentially recovered statistically: observers in different patches are equivalent in that they measure values of the cosmological parameters which fluctuate around the corresponding averages over the ensemble.

In conclusion, the inhomogeneous cosmologies constitute an appealing solution to the acceleration problem because of the economy in their hypothesis: no exotic form of energy or new theory of gravity is needed. Whether the acceleration is just apparent (Wiltshire 2007), (Cosmai et al. 2018) or effective (Lapi et al. 2023) is probably the next big question needing an answer and, in this respect, it is likely that the playground will be shifted to the other tensions afflicting the Λ-CDM, to discriminate among the 32different frameworks. As for the cosmological constant problem presented in the previous section, these models change the shape of the puzzle because they do not provide the renormalization conditions (4.9) and (4.11). Therefore, in a successful scenario one would still have to make sense of the possible backreaction of the vacuum energy, as it would happen in the modified gravity framework. In this respect, it appears undoubtful that a more fundamental understanding of the quantum fields on a dynamical background, possibly beyond perturbation theory is needed.

---

[64] That is, the system of the effective FLRW equations keeping into account the backreaction of the small scales matter inhomogeneities (Buchert 2000).

[65] Notice that the wall observer time is the actual cosmic time, since it is defined in the reference frame comoving (and hence clustering) with the matter elements.

[66] For example, a void observer would still see an isotropic CMB but could measure a different mean temperature and a different angular anisotropy scale (Wiltshire 2007).

## Acknowledgments

I thank O. F. Piattella for introducing me to this subject and for his precious comments on the manuscript. I am grateful to S. L. Cacciatori for his suggestions and for the stimulating discussions we had while conceiving this work. I also want to acknowledge that the discussion on the renormalization has been inspired by Cacciatori's quantum field theory lectures. I additionally thank G. Fava, M. Fontana and A. Massidda for their comments and many spontaneous exchanges. Finally, I thank A. Scali for pointing out the reference on the etymology of *chaos* in the ancient Greece.

## References

Abdalla E et al. (2022) Cosmology Intertwined: A Review of the Particle Physics, Astrophysics, and Cosmology Associated with the Cosmological Tensions and Anomalies. Journal of High Energy Astrophysics 34.

Akarsu O, Kumar S, Ozulker E, Vazquez J A (2021) Relaxing cosmological tensions with a sign switching cosmological constant. Physical Review D 104(12).

Alexandre B, Gielen S, Magueijo J (2024) Overall signature of the metric and the cosmological constant. Journal of Cosmology and Astroparticle Physics, 2024(02).

Aluri K et al (2023) Is the observable universe consistent with the cosmological principle? Classical and Quantum Gravity, 40(9).

Amendola L, Tsujikawa S (2010). Dark Energy: Theory and Observations. Cambridge University Press.

Anchordoqui L A, Antoniadis I, Lüst D (2024) Anti-de Sitter → de Sitter transition driven by Casimir forces and mitigating tensions in cosmological parameters. Physics Letters B, 855.

Balkenhol L et al (2023) Measurement of the CMB temperature power spectrum and constraints on cosmology from the spt-3g 2018 tt-te-ee dataset. Physical Review D, 108(2).

Bertone G, Hooper D (2018) History of dark matter. Reviews of Modern Physics, 90(4).

Bettini M (2015) Il grande racconto dei Miti classici. Il Mulino.

Bohr N (1913) On the constitution of atoms and molecules. The London, Edinburgh and Dublin Philosophical Magazine and Journal of Science, 26.

Bondi H, Gold T (1948) The Steady-State theory of the Expanding Universe. Monthly Notices of the Royal Astronomical Society 108.

Buchert T (2000) On Average Properties of Inhomogeneous Fluids in General Relativity I: Dust Cosmologies. General Relativity and Gravitation 32.

Cacciatori S L, Marrani A, Re F (2018) On generalized Lemaitre–Tolman–Bondi metric: Fractal matter at the end of matter–antimatter recombination. International Journal of Modern Physics D, 30 (11).

Capozziello S, Faraoni V (2010) Beyond Einstein Gravity: A Survey of Gravitational Theories

for Cosmology and Astrophysics. Fundamental Theories of Physics. Springer Netherlands.

Cartan E (1923) Sur les variétés à connexion affine et la théorie de la relativité généralisée (première partie). Annales scientifiques de l'École Normale Supérieure, 40.

Cartan E (1924) Sur les variétés à connexion affine et la théorie de la relativité généralisée (suite). Annales scientifiques de l'École Normale Supérieure, 41.

Coil A L (2013) The Large Scale Structure of the Universe. In Oswalt T D, Keel W C Plantes, Stars and Stellar systems. Springer.

Cottingham W N, Greenwood D A (2007) An introduction to the standard model of particle physics. Cambridge University Press.

Cosmai L, Fanizza G, Sylos Labini F, Pietronero L, Tedesco L (2018) Fractal universe and cosmic acceleration in a Lemaître–Tolman–Bondi scenario. Classical and Quantum Gravity 36 (4).

Chandrasekhar S (1995) Newton's Principia for the Common Reader. Oxford University Press.

Di Casola E, Liberati S, Sonego S (2015) Nonequivalence of the equivalence principles. American Journal of Physics 83.

Einstein A (1905a) Does the inertia of a body depend upon its energy-content? Annalen der Physik, 17.

Einstein A (1905b) On a heuristic point of view concerning the production and transformation of light. Annalen der Physik, 17.

Einstein A (1905c) On the electrodynamics of moving bodies. Annalen der Physik, 17.

Einstein A (1907) On the relativity principle and the conclusions drawn from it. Jahrbuch der Radioaktivität, 4.

Einstein A (1911) On the influence of gravitation on the propagtion of light. Annalen der Physik, 35.

Einstein A (1915a) Explanation of the perihelion motion of Mercury from the general theory of relativity. Preussische Akademie der Wissenschaften, Sitzungsberichte.

Einstein A (1915b) The field equations of gravitation. Preussische Akademie der Wissenschaften, Sitzungsberichte.

Einstein A (1915c) Fundamental ideas of the general theory of relativity and the application of this theory in astronomy. Preussische Akademie der Wissenschaften, Sitzungsberichte.

Einstein A (1915d) On the general theory of relativity. Preussische Akademie der Wissenschaften, Sitzungsberichte.

Einstein A (1916a) The foundations of the general theory of relativity. Annalen der Physik, 49.

Einstein A (1916b) Relativity: the special and the general theory. Methuen & Co Ltd.

Einstein A (1917) Cosmological considerations on the general theory of relativity. Sitzungsberichte der Preussischen Akadademie der Wissenschaften.

Einstein A (1923) Letter to Herman Weyl, 23 May 1923. From Princeton University website *The Digital Einstein Papers (*https://einsteinpapers.press.princeton.edu/).

Ellis G F R (2007) Issues in the Philosophy of Cosmology. In Butterfield J, Earman J Handbook of the philosophy of science: philosophy of physics. Elsevier.

Fizeau M H (1851) On the hypotheses relating to the luminous ether, and an experiment which appears to demonstrate that the motion of bodies alters the velocity with which light propagates itself in their interior. The London, Edinburgh, and Dublin Philosophical Magazine and Journal of Science, 2(14):568–573.

Fresnel A J (1816) Memoir on the diffraction of light, volume 1. Annales de Chimie et de Physique.

Friedman A (1922) On the curvature of space. Zeitschrift für Physik, 10:377–386.

Giulini D J W, Kiefer C, Lämmerzahl C (2003) Quantum Gravity: from Theory to Experimental

Search. Lecture Notes in Physics. Springer Berlin Heidelberg.
Green M B, Schwarz J H, Witten E (1987) Superstring theory, vol. 1-2. Cambridge University Press.
Heisenberg W (1925) On the quantum-theoretical reinterpretation of kinematical and mechanical relationships. Zeitschrift für Physik, 33.
't Hooft G (1980) Naturalness, chiral symmetry, and spontaneous chiral symmetry breaking. Nato Science Series B, 59:135–157.
Hoyle F (1948) A New Model for the Expanding Universe. Monthly Notices of the Royal Astronomical Society 108.
Hubble E (1929) A relation between distance and radial velocity among extra-galactic nebulae. Proceedings of the National Academy of Science, 15(3):168–173.
Hubble E (1934) The distribution of extra-galactic nebulae. The Astrophysical Journal, 79:8.
Kachru S, Kallosh, R, Linde A, Trivedi S P (2003) de Sitter vacua in string theory. Physical Review D 68 (4).
Konishi K, Paffuti G (2009) Quantum mechanics, a new introduction. Oxford University Press.
Kuhn T S (1957) The Copernican Revolution: Planetary Astronomy in the Development of Western Thought. Harvard University Press.
Kuhn T S (1962) The structure of scientific revolutions. University of Chicago Press.
Lane Z G, Seifert A, Ridden-Harper R, Wiltshire D L (2024) Cosmological foundations revisited with Pantheon+. Monthly Notices of the Royal Astronomical Society.
Lapi A, Boco L, Cueli M M, Haridasu B S, Ronconi T, Baccigalupi C, Danese L (2023) Little Ado about Everything: η-CDM, a Cosmological Model with Fluctuations-driven Acceleration at Late Times. The Astrophysical Journal, 959(2):83.
Lemaître G (1931) A homogeneous universe of constant mass and increasing radius accounting for the radial velocity of extra-galactic nebulae. Monthly Notices of the Royal Astronomical Society, 91:483-490.
Lorentz H A, Einstein A, Minkowski H, Weyl H (1923) The principle of relativity, a collection of original memoirs on the special and general theory of relativity. Dover Publications Inc.
Lorentz H A (1904) Electromagnetic phenomena in a system moving with any velocity smaller than that of light. Academy of Sciences of Amsterdam, 6.
Lüst D, Palti E, Vafa C (2019) AdS and the Swampland. Physics Letters B 797.
Mach E (1893) The science of mechanics. Open Court Publishing Co. (2nd ed.).
Macdonald A (1983) Clock synchronization, a universal light speed, and the terrestrial redshift experiment. American Journal of Physics, 51.
Maggiore M (2005) A Modern Introduction to Quantum Field Theory. Oxford University Press.
Mandelbrot B B (1983) The Fractal Geometry of Nature. W. H Freeman and Company.
Martin J (2012) Everything you always wanted to know about the cosmological constant problem (but were afraid to ask). Comptes Rendus Physique, 13(6-7):566-665.
Maxwell J C (1865) A dynamical theory of the electromagnetic field. Philosophical Transactions of the Royal Society of London, 155:459-512.
Michelson A A, Morley E W (1886) Influence of motion of the medium on the velocity of light. American Journal of Science, s3-31:377-385.
Minkowski H (1908) Space and Time. Address delivered at the 80th Assembly of German Natural and Physicians.
Misner C W, Thorne K S, Wheeler J A, Kaiser D I (2017) Gravitation. Princeton University Press.
Mukhanov V, Winitzki S (2007) Introduction to quantum effects in gravity. Cambridge University Press.

Nakahara M (2003) Geometry, topology and physics. Institute of Physics Publishing.

Newton I (1687) Philosophiae Naturalis Principia Mathematica. Londini, Iussu Societatis Regiae ac typis Josephi Streater.

Norton J D (1998) The cosmological woes of Newtonian gravitation theory. The Expanding Worlds of General Relativity. Birkhäuser Boston.

Obied G, Ooguri H, Spodyneiko L, Vafa C (2018) Se Sitter Space and the Swampland. arXiv:1806.08362.

Percacci R (2017) An introduction to covariant quantum gravity and asymptotic safety. World Scientific Publishing.

Perivolaropoulos L, Skar F (2022) Challanges for Λ-CDM: an update. New Astronomy Reviews, 95.

Perlmutter S et al (1999) Measurements of Omega and Lambda from 42 High-Redshift Supernovae. The Astrophysical Journal, 517(2):565-586.

Peskin M E, Schroeder D V (1995) An Introduction to Quantum Field Theory. Addison Wesley Publishing Company.

Peter P, Uzan J P (2009) Primordial Cosmology. Oxford University Press.

Piattella O (2018) Lectures Notes in Cosmology. Springer International Publishing.

Pietronero (1987) The fractal structure of the universe: correlations of galaxies and clusters and the average mass density. Physica A: Statistical Mechanics and its Applications 144.

Planck M (1901) On the law of distribution of energy in the normal spectrum. Annalen der Physik, 4.

Poisson E (2004) A Relativist's Toolkit: The Mathematics of Black-Hole Mechanics. Cambridge University Press.

Ribeiro (1992) On modeling a relativistic hierarchical (fractal) cosmology by Tolman's spacetime. I. Theory. The Astrophysical Journal 388.

Riess A G et al (2022) A comprehensive measurement of the local value of the Hubble constant with 1 kilometer per second per megaparsec uncertainty from the Hubble space telescope and the sh0es team. The Astrophysical Journal Letters, 934(1).

Ruede C, Straumann N (1997) On Newton-Cartan cosmology. Helvetica Physica Acta 70.

Schroedinger E (1926) An undulatory theory of the mechanics of atoms and molecules. Physical Review, 28.

Schwartz M D (2014) Quantum field theory and the standard model. Cambridge University Press.

Seeliger H (1895) Ueber das Newton'sche Gravitationsgesetz. Astronomische Nachrichten, 137.

Stachel J (1999) The Early History of Quantum Gravity (1916–1940). In: Iyer B R and Bhawal B (eds) Black Holes, Gravitational Radiation and the Universe. Fundamental Theories of Physics, 100. Springer.

Stephani H, Kramer D, Maccallum M, Hoenselaers C, Herlt E (2003) Exact Solutions of Einstein's Field Equations. Cambridge University Press.

Thiemann T (2007) Modern Canonical Quantum General Relativity. Oxford University Press.

Tung W (1985) Group Theory in Physics. World Scientific Publishing.

Wald R M (1984) General Relativity. The University of Chicago Press.

Wald R M (1994) Quantum field theory in curved spacetime and black hole thermodynamics. The University of Chicago Press.

Weinberg S (1972) Gravitation and Cosmology: principles and applications of the general theory of relativity. John Wiley and Sons.

Weinberg S (1989) The cosmological constant problem. Reviews of Modern Physics, 61:1–23.

Weinberg S (1995, 1996, 2013) The quantum theory of fields, vol. 1-3. Cambridge University Press.

Weinberg S (2008) Cosmology. Oxford University Press.

Wiltshire D L (2007) Cosmic clocks, cosmic variance and cosmic averages. New Journal of Physics 9 (10).

Whittaker E (1989) A History of the Theories of Ether and Electricity: Vol. I: The Classical Theories; Vol. II: The Modern Theories, 1900-1926. In A History of the Theories of Ether & Electricity. Dover Publications.

Will C M (2014) The Confrontation between General Relativity and Experiment. Living Reviews in Relativity, 17(1).

Young T (1804) The Bakerian Lecture. Experiments and calculations relative to physical optics. The Philosophical Transactions of the Royal Society of London, 94:1–16.

Zinn-Justin J (1996) Quantum field theory and critical phenomena. Oxford University Press.

Zel'dovich Ya B, Krasinski A (1968) The cosmological constant and the theory of elementary particles. Soviet Phyics Uspekhi, 11:381-393.

**Names Index:** Einstein Albert**,** 't Hooft Gerard**,** Newton Isaac**,** Weinberg Steven

**Subject Index:** Accelerated expansion, Cosmological constant, Expansion of the universe, General Relativity, Inhomogeneous cosmologies, Modified gravity, Naturalness, Newtonian Cosmology, Newtonian Gravity, Quantum fields, Quantum fluctuations, Redshift, Relativistic cosmology, Renormalization, Self-adjusting mechanism, Semiclassical Einstein equations, Semiclassical gravity, Standard model, String theory, Supernovae, Supersymmetry, Vacuum energy, Vacuum expectation value, Zero-point energy.